\documentclass[apj]{emulateapj}
\usepackage{amsmath,amssymb,graphicx,color,longtable,natbib,multirow}

\newcommand{\simlt}
{\ifmmode
{ \raisebox{-.4em}{$<$}\atop\sim}
\else
{$\raisebox{-.4em}{$<$}\atop\sim$}
\fi}

\shorttitle{The Theoretical Astrophysical Observatory}
\shortauthors{Bernyk et al.}

\begin{document}
\normalsize

\title{The Theoretical Astrophysical Observatory\lowercase{\footnotemark[$\dagger$]}: Cloud-Based Mock Galaxy Catalogues}

\author{Maksym Bernyk\altaffilmark{1},  Darren J. Croton\altaffilmark{1} Chiara Tonini\altaffilmark{2,1}, 
Luke Hodkinson\altaffilmark{1}, Amr H. Hassan\altaffilmark{1}, Thibault Garel\altaffilmark{3,1}, 
Alan R. Duffy\altaffilmark{1}, Simon J. Mutch\altaffilmark{2,1}, Gregory B. Poole\altaffilmark{2,1}, 
Sarah Hegarty\altaffilmark{1}}

\affil{$^1$Centre for Astrophysics \& Supercomputing, Swinburne University of Technology, PO Box 218, Hawthorn, Victoria, 3122, Australia}
\affil{$^2$School of Physics, University of Melbourne, Parkville, Victoria 3010, Australia}
\affil{$^3$Centre de Recherche Astrophysique de Lyon, Universit\'{e} de Lyon, Universit\'{e} Lyon 1, CNRS, Observatoire de Lyon, ~9 avenue Charles Andr\'{e}, 69561 Saint-Genis Laval Cedex, France}

\begin{abstract}
We introduce the Theoretical Astrophysical Observatory (TAO), an online virtual laboratory 
that houses mock observations of galaxy survey data. Such mocks have become an integral 
part of the modern analysis pipeline. However, building them requires an expert knowledge 
of galaxy modelling and simulation techniques, significant investment in software 
development, and access to high performance computing. These requirements make it 
difficult for a small research team or individual to quickly build a mock catalogue 
suited to their needs. To address this TAO offers access to multiple cosmological 
simulations and semi-analytic galaxy formation models from an intuitive and clean 
web interface. Results can be funnelled through science modules and sent to a dedicated 
supercomputer for further processing and manipulation. These modules include the ability 
to (1) construct custom observer light-cones from the simulation data cubes; (2) generate 
the stellar emission from star formation histories, apply dust extinction, and compute 
absolute and/or apparent magnitudes; and (3) produce mock images of the sky. All of 
TAO's features can be accessed without any programming requirements. The modular 
nature of TAO opens it up for further expansion in the future.
\end{abstract}

\keywords{galaxy formation, mock catalogue, light cone, online tools, cloud computing}

\footnotetext[$\dagger$]{https://tao.asvo.org.au/}

\maketitle

\section{Introduction}
\setcounter{footnote}{0}

Astronomy has entered an era of survey science, thanks almost solely to 
instruments and telescopes that can sample unprecedented volumes of
the cosmos with unparalleled sensitivity out to great distances. Riding this
wave, the field of galaxy formation and evolution has grown to become one of the
most active research areas in astrophysics. Progress feeds off progress, where
increasingly detailed observations of galaxies are used to build new theories,
that in turn lead to predictions, which can direct and be tested against new
observations. As the instruments have increased in sophistication, so have the amount
of data they collect. Similarly, simulations of galaxies and the Universe
have grown to keep pace.

It is thus not surprising that data access has become a signature
of this new era. Internet and cloud technologies allow scientists to store
and retrieve large scientific datasets remotely. This is sometimes
necessary since data volume and complexity often require resources beyond 
what is locally available. But even when not necessary it is frequently 
desired, as relocating storage and processing off-site reduces overheads, 
simplifies data management, and facilitates data sharing. 

Two notable examples are the Sloan Digital Sky Survey
\citep[SDSS, ][]{Abazajian2003} ``SkyServer'', which hosts imaging, spectra,
spectroscopic and photometric data; and the German Astrophysical Virtual
Observatory \citep[GAVO, ][]{Lemson2006}, which houses the Millennium Simulation
theoretical data products. Both on-line repositories are accessible by means of
the Structured Query Language (SQL), and due to their accessibility, both have
vastly increased the scientific value of the data they hold through data re-use.
There are many ways this benefit can be measured, arguably the most important being
an increased number of scientific publications. Such publications come
predominantly from researchers who had nothing to do with the original data
production.

In large part due to this ease of access, the division between observer and
theorist has faded somewhat. Observers now routinely use cutting edge
theoretical models in their analysis, and theorists compare model predictions 
against observational data. This has all meant that the modern astronomer now
routinely works across many traditional boundaries, often combining multiple
disparate data products to undertake their science.

The focus of the present work is on access to theoretical survey data, such as
cosmological-scale dark matter simulations and galaxy formation models. Many groups around the
world are currently producing state-of-the-art theory products whose value to
the community is immense. However access from outside the group is often prohibitive, even when the
authors are happy for others to use their work. Furthermore, comparing different
simulations and models on an equal footing can be extremely problematic due to data
size and transport barriers, data format differences, and complexity. This makes
understanding how to correctly use the data challenging for the non-expert.

Access is but one part of the puzzle however. To be compared fairly to
observations, simulations must typically be modified to look more like the data
being compared against. This can include: mapping the simulation cube into an
observed light-cone where distance also equates to time evolution in the
simulation, calculating absolute and apparent magnitudes for model galaxies in
select filters from their star formation and metallicity histories, and
``observing'' mock galaxies to generate images similar to that which would be
collected by a CCD. All are non-trivial tasks that require great care to implement correctly.

Some effort has already gone in to producing such tools for the community. For example,
the Mock Map Facility \citep[MoMAF, ][]{Blaizot2005} allows a user to build mock
galaxy catalogues using the GALICS semi-analytic model \citep{Blaizot2004}.
In a similar fashion, the Millennium Run Observatory \citep{MRO2012}
provides a powerful set of tools to access and visualise mock catalogues based 
on the Millennium Run suite of dark
matter simulations \citep{Springel2005GADGET}.

In this paper we present a new online tool -- the Theoretical Astrophysical
Observatory (TAO) -- that aims to further address the problem of community data access and,
more specifically, simplifies the process of building mock galaxy catalogues to more individualised
specifications. 
This paper is structured as follows: An introduction to TAO is presented in 
Section~\ref{TAO}. In the subsequent sections we then describe the first 
four TAO science modules: the basic
galaxy and simulation selection tools (Section~\ref{galaxy-module}), the
light-cone module (Section~\ref{Lightcones}), the spectral energy
distribution (SED) module (Section~\ref{SEDs}), and the mock image generation 
module (Section~\ref{mockimage}). We then explore usage cases in
Section~\ref{applications} that demonstrate the utility and functionality of TAO.
Section~\ref{summary} concludes with a summary.

For all results presented the cosmology of the simulation from which the result
was drawn is assumed unless otherwise indicated, and we refer the reader to the 
associated reference for further details.

\section{The Theoretical Astrophysical Observatory - An Overview} \label{TAO}

The Theoretical Astrophysical Observatory (TAO) provides web
access to cloud-based\footnote{In this paper we define the 
``cloud'' as Infrastructure as a Service (IaaS), which includes data storage and
access, software infrastructure, and the use of supercomputer facilities.} mock 
extragalactic survey data, generated using sophisticated
semi-analytic galaxy formation models that are coupled to large N-body
cosmological simulations. TAO is designed to be flexible, so that different
simulations and models can be stored and accessed from a single location  
with a consistent data format. The interface for TAO is clean and built with 
simplicity in mind. All of TAO's features require no programming knowledge
to use, maximising accessibility to astronomers, be they observers or theorists.

A major feature of TAO is its ability to post-process the hosted data 
for different scientific applications. This is  achieved through a number of science modules
that can be chained in user-specified configurations, depending on the desired
requirements of the astronomer and the module functionality. In addition, this
modular design makes TAO readily expandable with new functionality in the
future.

\begin{itemize}
\item \textit{Simulation data module.} This core module provides direct
access to the original simulation and semi-analytic galaxy formation model
data stored in the TAO SQL database. The user can specify the desired galaxy and
dark matter halo properties to be retrieved at an epoch of interest from the
simulation box (see Section~\ref{galaxy-module}).
\item \textit{Light-cone module.} This module remaps the spatial and temporal
distribution of galaxies in the original simulation box on to that of the
observer light-cone. The parameters of the cone are user configurable (see
Section~\ref{Lightcones}).
\item \textit{Spectral energy distribution module (SED).} This module retrieves
the star formation and metallicity histories for each galaxy (either in the box
or cone) from the TAO database and applies a user-selected stellar population
synthesis model and dust model to produce individual galaxy spectra. These
spectra are convolved with a set of filters to compute both apparent
and absolute magnitudes (see Section~\ref{SEDs}).
\item \textit{Image module.} This module takes the output of both the
light-cone and SED modules to construct user defined mock images. Images can be
customised using a range of properties, such as sky area, depth, and observed
filter (see Section~\ref{mockimage}).
\end{itemize}

\begin{figure*}
  \begin{center}
  \centering
  \includegraphics[scale=0.4]{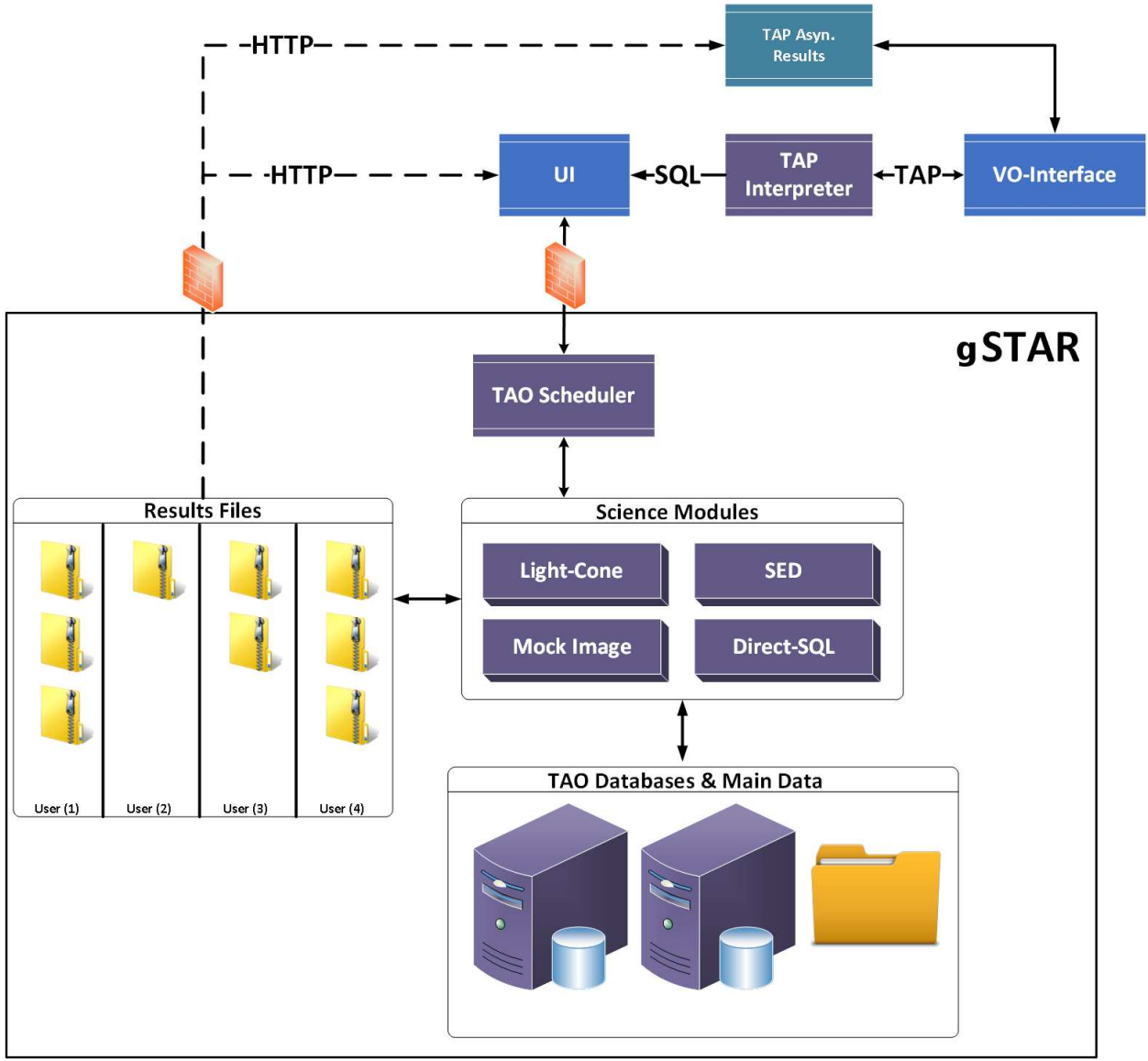}
  \caption{A TAO architecture diagram. The main TAO database, science modules, and results 
	storage space are located on the gSTAR supercomputer at Swinburne University. The user 
	interface and Table Access Protocol (TAP) server are hosted on a separate web server 
	and accept job requests. Jobs are queued on the supercomputer via the jobs scheduler. 
	Upon the job completion results are available for download over the internet.
}
  \label{tao-arch}
  \end{center}
\end{figure*}

In Figure~\ref{tao-arch} we show a broad overview of the TAO
infrastructure. At the top level we define the connection between the account
based user interface and user database. In the middle level the various science
modules are shown, as introduced above and which will be described in more
detail in the subsequent sections. Both public and private data storage,
containing the dark matter simulations, galaxy data, photometry etc., are shown
in the lower level. 
On the back-end TAO is supported by a scalable database cluster hosted on the gSTAR 
supercomputer\footnote{http://www.astronomy.swin.edu.au/supercomputing/} at the
Swinburne University of Technology. Each component of TAO affects the user
experience and workflow, speed of mock data generation and retrieval, and the
quality and utility of the final mock catalogue.

Users interact with TAO through a simple web form, where they can select a dark
matter simulation, galaxy formation model, a box or cone geometry, and the
associated parameters which define each. The desired galaxy and simulation
properties to be included in the mock, including both absolute and apparent
magnitudes in various filters, are all specified by the user. These selections are then 
passed to the back-end science modules via an XML parameter file, which can 
also be retrieved from the web interface for
reference or later resubmission. Any required computations, e.g. to build a
light-cone or set of SEDs, are then triggered on the gSTAR supercomputer for
processing. The user is notified via email when their mock catalogue is 
completed, which may take from minutes to many hours depending on the size
of the task. TAO offers a choice of output formats, including CSV, HDF5 
and FITS. The final mock catalogue can then be downloaded directly from the
TAO website ``History'' tab to the user's local machine. More advanced 
SQL/ADQL querying of the data is accessible through a VO Table Access Protocol 
(TAP) client, such as TOPCat.

\begin{figure*}
  \begin{center}
  \centering
  \includegraphics[scale=0.52]{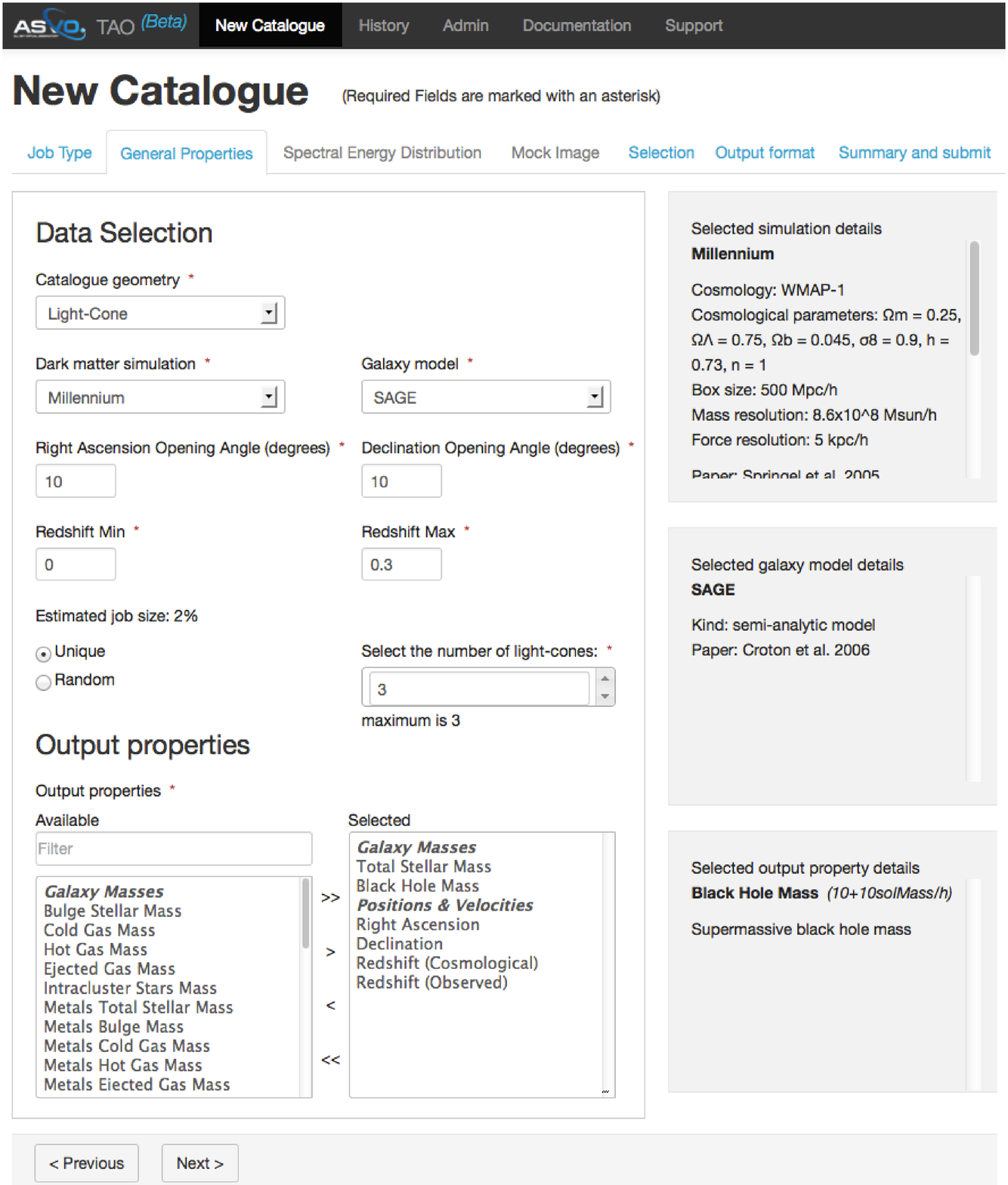}
  \caption{The TAO web interface, showing how complex queries can be generated through 
	a simple selection of galaxy and simulation properties, plus additional options to 
	access the various science modules.}
  \label{tao-screenshot}
  \end{center}
\end{figure*}

In Figure~\ref{tao-screenshot} we show the TAO web interface, highlighting its
minimalist yet functional design. The user interface design objective is 
to provide a simple portal that makes using the science modules easy and intuitive. 
In the spirit of reaching as many astronomers as possible, no programming
knowledge (SQL or other) is required to use any part of TAO, keeping the barrier for
access low.

TAO is part of the larger All-Sky Virtual Observatory (ASVO) project\footnote{http://www.asvo.org.au/}, whose goal is to 
federate astronomy data and serve this to the wider community via the cloud. 
The ASVO constitutes a major infrastructure investment 
that links observational data with theoretical capabilities. It establishes a 
platform from which astronomers can optimally access and exploit the exponential
growth in astronomical data volume in the coming decade. As a first release, the
ASVO will include cloud access to the SkyMapper data archives \citep{Keller2007}, 
in addition to the simulated data provided through TAO. Ultimately it is expected 
that the ASVO will incorporate data at multiple wavelengths, including radio 
observations from the upcoming Australian Square Kilometre Array Pathfinder 
(ASKAP) telescope \citep{Johnston2008}.

\section{The Galaxy and Simulation Module} \label{galaxy-module}

Simulated galaxy data is the core product of TAO. Hence, it is useful to review
the basics of how TAO data is generated and the properties that define it. We will
do this by considering dark matter simulations and semi-analytic galaxy modelling
in turn. 
More detail on these methods can be found in \citet{Baugh2006,Croton2006,Benson2012}. 
We also clarify some of the technical requirements of TAO to hold and use this data. 
These include the requisite data format and minimum galaxy and halo properties 
required by the core and higher-level science modules. 
An up-to-date list of the data hosted can be found at the TAO 
website\footnote{https://wiki.asvo.org.au/display/TAOC/Available+Data+Sets}.

\subsection{Dark matter simulations and large-scale structure}
\label{sims}

An efficient way to simulate the universe inside a supercomputer is
to focus on the dominant mass distribution and its evolution. 
This usually involves running a collisionless N-body simulation 
in a volume that is large enough to be representative of the Universe 
as a whole, and provides a significant reduction
in computational effort at fixed resolution compared with hydrodynamic 
galaxy formation simulations. Hydrodynamic effects are
complicated and slow to compute numerically relative to the rather simple
calculations required in a gravity-only simulation. Hence, gas and galaxies are
often added later in post-processing using semi-analytic or other statistical
techniques (see below).

As the universe evolves gravity pulls small structures together to assemble
larger structures (i.e. hierarchical growth). Within the numerical simulation,
such ``halos'' are typically identified using a Friends-of-Friends (FoF)
algorithm \citep{Davis1985, Springel2001, More2011}, which detects
gravitationally bound systems of particles and determines their
properties. Structures within structures (i.e. sub-structures) can be
found using a variety of methods \citep[e.g.][]{Springel2005, Behroozi2011}.
Such sub-structures are typically expected to host the smaller satellite
galaxies and are subservient to the larger halo and central galaxy at the halo
centre.

\begin{figure*}
  \begin{center}
  \begin{minipage}{175mm}
  \centering
  \includegraphics[scale=0.55]{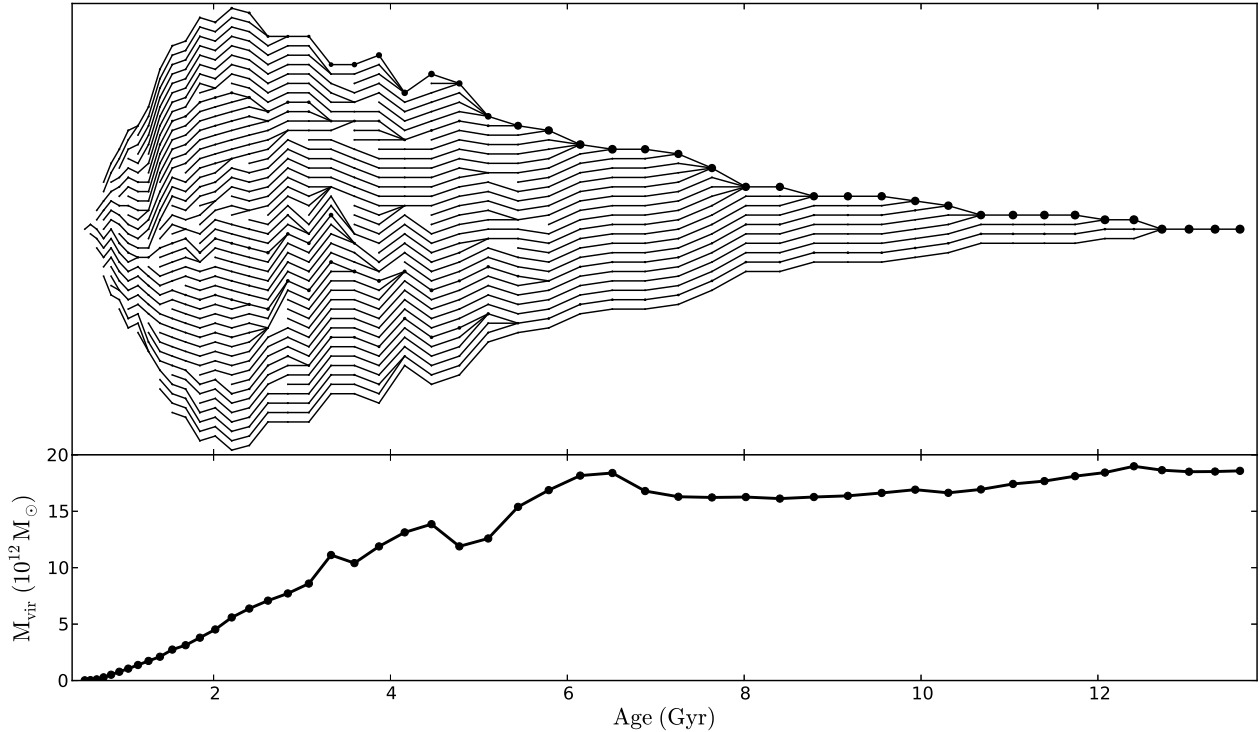}
  \caption{A dark matter halo merger tree drawn from the Millennium Simulation
  (top) and its mass evolution (bottom) with time. This halo has a final mass of
  $1.9 \times 10^{13}\, {\rm M_\odot}$ (assuming $h=0.73$) and would be typical
  of a group sized system in the real Universe.}
  \label{merger-tree}
  \end{minipage}
  \end{center}
\end{figure*}

This information, calculated across all time-steps in a simulation for a
particular object, defines its merger tree. The collection of such trees is 
then used as input to construct a galaxy formation model. 
An example halo merger tree from the Millennium Simulation \citep{Springel2005}
is shown in Figure~\ref{merger-tree}. Here, the top panel shows the tree itself
for a $1.9 \times 10^{13}\, {\rm M_{\odot}}$ halo at $z=0$ (assuming $h=0.73$),
while the corresponding mass growth history with time is shown in the lower panel. 

Simulations are run for different science goals, and each allow the exploration 
of different physics depending on how they were set-up. TAO provides 
the user with a choice of dark matter simulations, run with various sizes, mass
resolutions and cosmological parameters.

\subsection{Modelling the evolution of galaxies}

There are a number of ways to model the evolution of a galaxy inside a dark
matter halo. For the higher-level science modules, TAO requires the galaxies in
its database to have a minimum set of properties. As discussed below in
Section~\ref{input-data}, each galaxy record in the TAO database should contain
information about its position, merger history, and star formation history. So,
at a minimum, the methodology used to generate the galaxy population must
produce this information.

The best suited methodology for this purpose is that of semi-analytics \citep{White1991}.
Semi-analytic models not only calculate all the required properties for
galaxies, but also link their evolution over time using the halo merger trees
to follow the growth histories. 
Note that any other method that produces the minimum set of properties is also acceptable; we focus on
semi-analytics here as it is the most common and nicely illustrates the
requirements of the TAO system.

A semi-analytic galaxy model takes as input a dark matter halo merger tree
and evolves its baryonic content with time using prescriptions that describe 
the phenomenology of each key galaxy formation process.
\begin{enumerate}
\item As a dark matter halo grows its potential well gets deeper and hence
attracts baryons in the form of diffuse gas from the surrounding medium.
\item This gas cools, conserving angular momentum as it falls to
the centre of the halo and forms a rotationally supported disk.
\item Within this flattened disk stars begin to form. A galaxy is born.
\item Each episode of star formation results in a distribution of stellar
masses sampled from the initial mass function.
\item The most massive stars are short-lived and explode as supernovae,
injecting metals and energy back into the interstellar and intergalactic medium.
\item Galaxies merge as hierarchical growth proceeds,
resulting in morphological evolution and the birth of super-massive black holes.
\item Supernova and super-massive black hole feedback can heat and/or remove gas from the
disk and halo, gas that would otherwise contribute to the formation of the next
generation of stars.
\item Thus, an equilibrium state is established as the galaxy goes through
a cycle of stellar birth, feedback, gas heating/removal and star formation suppression.
\end{enumerate}

The above processes provide us with a history of galaxy
properties. The simulations and models
used in TAO are often large, tracking many tens of millions of halos and hence
galaxies. It is from the properties contained in such large catalogues
that the output of TAO is derived.

\subsection{TAO data characteristics} \label{input-data}

Each simulation and galaxy model will potentially contain a vast array of properties 
of interest to astronomers. TAO groups its data for each halo--galaxy pair into the following categories:
\begin{itemize}
  	\item Baryonic masses, such as stellar mass, cold gas mass, and hot halo gas. Metals for each baryonic 
		component are included in this category.
  	\item Other galaxy properties, such as the star formation rate, disk scale radius, and cooling and AGN heating rates.
  	\item Halo properties, such as virial mass, radius and velocity, the halo spin vector, and velocity dispersion.
  	\item Co-moving positions and physical (peculiar) velocities common to both halos and galaxies. These are always given in the Cartesian coordinate system of the simulation box.
  	\item Additional simulation properties, including the temporal snapshot number and any galaxy and halo IDs.
\end{itemize}
For those considering importing their data into TAO, in Table~\ref{requirements} we provide a summary 
of the minimum property requirements that TAO needs to operate, broken down by science module.
\newline \newline \noindent \textbf{Light-Cone Requirements:} To build a mock light-cone the 
data must contain Cartesian coordinates for
each halo--galaxy pair from the original simulation box; these are then converted
to angular coordinates and redshifts using methods described in Section
\ref{Lightcones}. Furthermore, conversion to redshift space requires Cartesian
velocities (i.e. proper motion of the galaxies).
\newline \newline \noindent \textbf{SED Requirements:} TAO must be able to walk each halo merger tree
from any point in its history. Hence, for each object we require pointers that
identify both the previous time-step progenitor and that link the sequence of
subhalos in a FoF halo at a given time-step in order of decreasing mass. We
adopt the SUBFIND system of pointers, assuming they are stored in depth-first
order for each merger tree, as illustrated by Figure~11 of the Supplemental Material 
in \cite{Springel2005}. 
Furthermore, to calculate magnitudes the SED module must be able to extract star formation
and metallicity histories for each galaxy. Also, the initial mass function
(IMF) assumed when calibrating the model is required (e.g., this typically constrains the  
recycling fraction of mass returned to the interstellar medium from stellar winds).
\newline \newline \noindent \textbf{Mock Image Requirements:} The image module requires some measure of morphology 
to correctly construct the right shape for each galaxy before rendering. This may take the 
form of seperate disk and bulge stellar masses, for example.

\begin{table}
\centering
\begin{tabular}{|l|l|}
\hline
\parbox[c][2em][c]{0.08\textwidth}{Module} & 
\parbox[c][2em][c]{0.27\textwidth}{Minimum information required}
\\
\hline
\hline
\parbox[c][2.5em][c]{0.08\textwidth}{
\textbf{Simulation \& Model (\S\ref{galaxy-module})}}
& \parbox[c][4.5em][c]{0.27\textwidth}{
$\bullet$ A list of the simulation snapshot time-steps. The box size,
particle mass resolution and assumed cosmological parameters.} \\
\hline
\multirow{4}{*}{\textbf{Light-Cone (\S\ref{Lightcones})}}
& \parbox[c][4.5em][c]{0.27\textwidth}{
$\bullet$ Co-moving coordinates ($x,\ y,\ z$) and physical (peculiar) velocities ($v_x,\ v_y,\ v_z$)
for each halo/galaxy at each time-step.} \\
& \parbox[c][3.5em][c]{0.27\textwidth}{
$\bullet$ IDs that associate subhalos/satellites with their parent central halo/galaxy.} \\
\hline
\multirow{5}{*}{\textbf{SED (\S\ref{SEDs})}}
& \parbox[c][4.5em][c]{0.27\textwidth}{
$\bullet$ Indices connecting the history of each halo/galaxy across time through the galaxy merger tree (see Section~\ref{input-data}).} \\
& \parbox[c][2.5em][c]{0.27\textwidth}{
$\bullet$ Galaxy star formation rate and metallicity at each time-step.} \\
& \parbox[c][2.5em][c]{0.27\textwidth}{
$\bullet$ The initial mass function assumed by the galaxy model.} \\
\hline
\multirow{3}{*}{\textbf{Mock Images (\S\ref{mockimage})}}
& \parbox[c][3.5em][c]{0.27\textwidth}{
$\bullet$ A measure of galaxy morphology, such as independent galaxy disk 
and bulge stellar masses.} \\
& \parbox[c][2.5em][c]{0.27\textwidth}{
$\bullet$ The output from the light-cone and the SED modules.} \\
\hline
\end{tabular}
\caption{Data requirements for the TAO modules.}
\label{requirements}
\end{table}

\subsubsection{Units}

We conclude this section by defining the unit and Hubble constant conventions 
assumed in TAO. The use of little $h$ in particular can differ significantly 
between the theory and observational communities (and even within a community) 
and care must always be taken \citep{Croton2013}. Unless otherwise stated: 
\begin{itemize}
	\item Masses adopt $h^{-1} \rm M_{\odot}$.
	\item Positions adopt $h^{-1} \rm Mpc$ and are in co-moving coordinates.
	\item Sizes adopt $h^{-1} \rm Mpc$ and are in physical coordinates.
	\item Velocities adopt $\rm km \, s^{-1}$ and are in physical coordinates.
	\item Star formation rates adopt $\rm M_{\odot} \, yr^{-1}$.
	\item Heating and cooling rates are in units of $\log_{10} \rm erg \, s^{-1}$.
	\item Photometric magnitudes are calculated assuming the $h$ value of the dark matter 
	simulation used to produce the mock catalogue.
\end{itemize}

\section{The Light-Cone Module} \label{Lightcones}

N-body simulations of the type discussed in the previous section typically
adopt a periodic three-dimensional box geometry. This is in contrast to
the geometry observed with a telescope, where galaxies are seen strung out
along the observer's light-cone. Converting between a box and a light-cone for
the purpose of building a mock catalogue is a mechanical yet non-trivial task
\citep{Blaizot2005,Kitzbichler2007,Carlson2010}. In TAO, this is taken care of 
by the light-cone module. 

The module can work in two modes. The first is to produce light-cones from entirely 
unique structure, meaning any galaxy can be selected for the cone from any point in 
its history at most once. This mode is similar to the method described by \cite{Carlson2010}. 
In our work we concentrate on automated light-cone construction using an analytic 
solution for the inclination angle calculations.

The second mode is for larger light-cones. It is used when it is not possible to fill 
the light-cone volume without repetitions of the structure. This mode is similar 
to that described in \cite{Blaizot2005}. Our algorithm, however, additionally involves random mirroring of 
the simulation box along the principal axis, increasing the number of pseudo unique 
volumes by a factor of three. Also, to make the randomised structure even more unique a translation
is performed, random shifting along the principal axis. We describe both of these enhancements below.

To explain cone construction, we start with some basic concepts and build up 
to the more sophisticated method used by TAO.

\subsection{Basic light-cone construction}

To build a basic light-cone we place an observer at one of the corners of the
simulation box and have them ``look out'' at the model galaxy distribution. We
do this by remapping the Cartesian coordinates of each galaxy into their
angular positions in right ascension (RA), declination (Dec) and radial distance ($d$).
This operation defines the basic cone geometry in real-space.
\begin{equation} \label{basic_cone}
\begin{cases}
  d = \sqrt{ x^{2} + y^{2} + z^{2} } \\
  {\rm RA}  = \arctan \left( \frac{y}{x} \right) \\
  {\rm Dec} = \arcsin \left( \frac{z}{d} \right)
\end{cases}
\end{equation}
Here $x$, $y$, and $z$ are the co-moving coordinates along the principal axes of the simulation box.

For TAO to provide a redshift for each galaxy on the cone it is necessary to 
invert the radial distance, $d$, using the
distance--redshift relation for the given simulation cosmology \citep{Hogg1999}.
This is defined by
\begin{equation}
d = d_{H}\,\int_0^z\frac{dz'}{E(z')}~,
\label{eqn:dzrelation}
\end{equation}
where $d_{H} = c/{\rm H_{0}}$ is the Hubble distance, $c$ is the speed of light, 
${\rm H_{0}}$ the Hubble constant,
$z$ is now redshift, and the expansion factor $E(z)$ is defined as
\begin{equation}
E(z) \equiv \sqrt{\Omega_{M} (1 + z)^{3} + \Omega_{k} (1 + z)^{2} + 
\Omega_{\Lambda}}~.
\end{equation}
$\Omega_{M}$, $\Omega_{k}$, and $\Omega_{\Lambda}$ are the
matter density, curvature, and cosmological constants, respectively.
To obtain the real-space (i.e. cosmological) redshift $z$, TAO simply inverts 
Equation~\ref{eqn:dzrelation}.

However TAO also has the capacity to generate cones in redshift-space, meaning with 
line-of-sight peculiar velocities factored into the radial distance. To determine 
the redshift-space (i.e. observed) redshift for each galaxy, $z_{\rm obs}$, we solve
\begin{equation}
 (1+z_{\rm obs}) = (1+z) (1+z_{\rm pec})~,
\label{eqn:cosmological_z}
\end{equation}
where $z$ is the real-space redshift from above, and $z_{\rm pec} = v_{\rm pec} / c$ is 
the redshift distortion due to the objects peculiar velocity $v_{\rm pec}$, given by
\begin{equation}
  v_{\rm pec} = \left( \frac{x}{d} v_x + \frac{y}{d} v_y + \frac{z}{d} v_z \right) ~.
\end{equation}
Here $v_x$, $v_y$, and $v_z$ are the physical coordinates of the velocity for the galaxy in
$\rm km/s$. As such, each TAO galaxy returns a cosmological redshift (real-space, 
from Equation~\ref{eqn:dzrelation}) and an observed redshift (redshift-space, 
from Equation~\ref{eqn:cosmological_z}), based on its position and dynamics in the cone.

\begin{figure*}
	\begin{center}
		\begin{minipage}{160mm}
			\centering
			\includegraphics[scale=0.65]{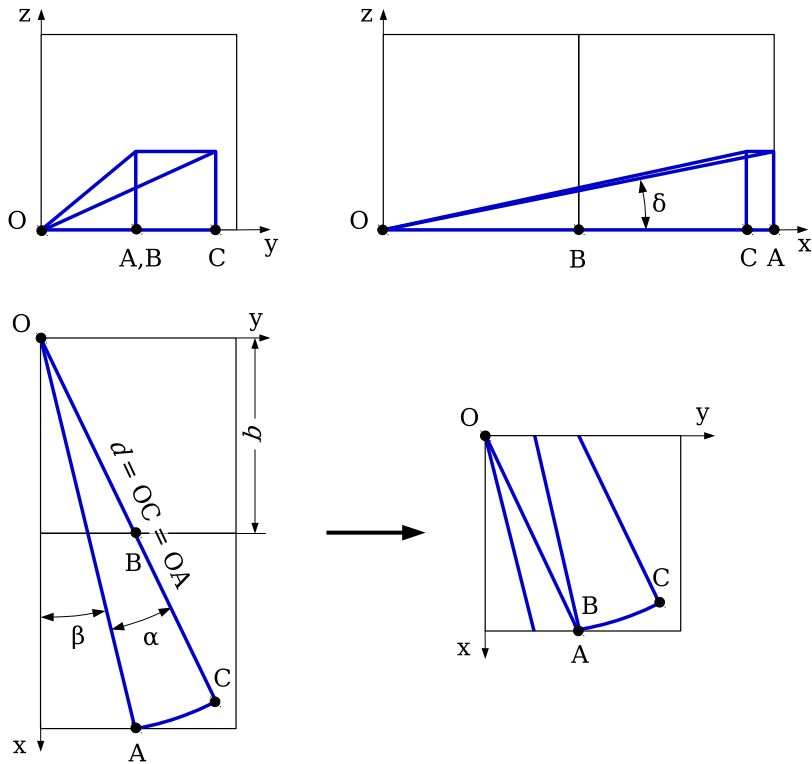}
			\caption{The geometry related to finding a unique light-cone path through a replicated simulation cube with side-length $b$, projected in each Cartesian plane (upper two and lower-left panels). The user-selected cone opening angles are marked by $\alpha$ for RA and $\delta$ for Dec. $\beta$ is the RA translation TAO must find to ensure the cone volume never overlaps as it extends a distance $d$ through the boxes. A, B and C mark points in the last replicated simulation box used in Equation~\ref{condition2}. Once a non-overlapping (i.e. unique) path is found, the boxes are collapsed back to a single box (lower-right panel) and the entire simulation is rotated so-as to point the cone at the requested area of the sky. See Section~\ref{unique} for a full description.}
			\label{unique}
		\end{minipage}
	\end{center}
\end{figure*}

\subsection{Expanding the cone beyond the box}

A problem with the above cone construction becomes apparent when building cones
that are deeper in radial extent than the box from which the cone is cut.
However, there are a number of ways to deal with this. All of them rely on the
fact that most modern simulations are run assuming periodic boundary
conditions, meaning that each side of the box connects seamlessly with its
opposite side. Periodic boundary conditions are a requirement for all simulation data in TAO. 
There are now two cases to consider.

\subsubsection{Unique cones}
\label{unique}

\begin{figure*}
	\begin{center}
		\begin{minipage}{160mm}
			\centering
			\includegraphics[scale=0.3]{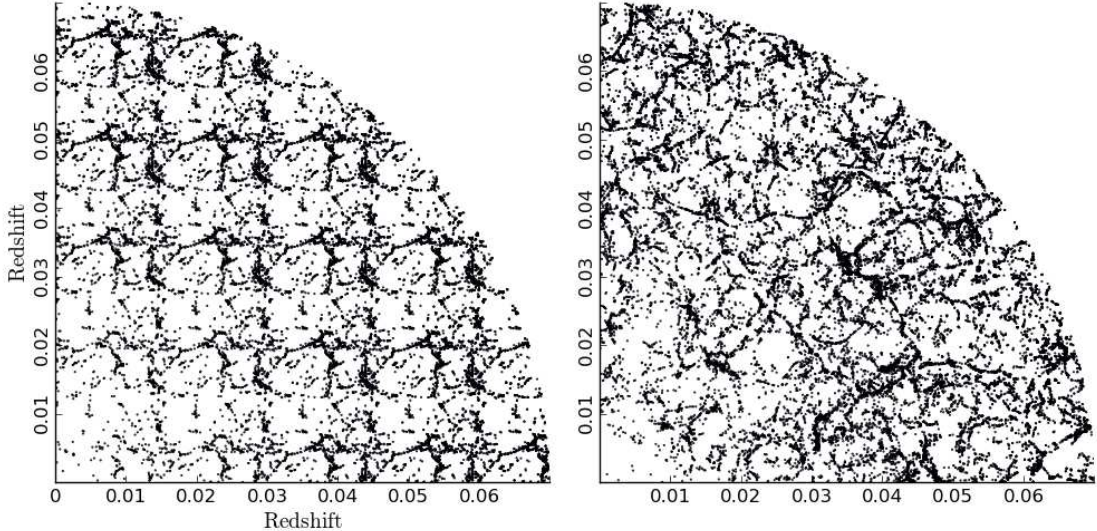}
			\caption{Two volume limited light-cones built from a simulation with a box
				side-length of 62.5 $h^{-1}\, {\rm Mpc}$. The left cone shows the result of
				standard box replication, while the right cone includes the random
				rotation, shifting and mirroring techniques employed by TAO to minimise any
				periodicity, as discussed in Section~\ref{random}.}
			\label{replication}
		\end{minipage}
	\end{center}
\end{figure*}

To construct a unique extended cone we replicate the box in the desired direction and
continue the cone construction into this `new' box. To ensure the cone is
unique -- i.e. any given galaxy in the box along any point in its history is only 
featured in the cone at most once -- we carefully select the angle the cone 
initially cuts through the box so that it never overlaps with itself as it extends. 
A similar method is used in \cite{Carlson2010}. Here we present an analytical 
method to find the optimal (or at least sufficient) unique path.

In Figure~\ref{unique} we present three projections of an extended box 
and the desired unique volume through it. Note that initially we ignore 
the absolute direction in RA and Dec the user has chosen for their unique 
cone; later we will rotate the entire simulation appropriately so that 
the found unique volume points at the requested patch of sky, correctly 
accounting for declination effects. 

The angles $\alpha$ and $\delta$ in Figure~\ref{unique} represent the cone 
opening angles in the RA and Dec planes respectively, taken from the 
minimum and maximum angles requested by the user. To find the unique volume 
we need to determine the angle $\beta$ (confined to the RA plane), which 
is the required cone offset such that, as the cone extends back through the replicated 
boxes, there are no intersections with any previous volume, as illustrated.

First, the following two minimal conditions must be met:
\begin{equation}
\begin{split}
d \, \sin( \alpha + \beta ) \leq b~,\\
d \, \sin( \delta ) \leq b~.
\end{split}
\label{condition1}
\end{equation}
In other words, at its furthest end, $d$ (from Equation~\ref{basic_cone}), 
the opening width of the cone in both 
directions cannot be larger than the width of the simulation box, $b$. 
Otherwise we may be extending into volume that the method cannot guarantee is unique.

Second, considering the lower-left panel in Figure~\ref{unique} which 
looks down onto the RA plane, the length along the y-axis from 0 to B
must be shorter than or equal to the length along the y-axis from 0 to A, as marked. 
This requirement ensures that, when the cone path through the replicated 
boxes is collapsed back to a single box, as shown in the lower-right panel, 
that point A does not lie inside the earlier part of the cone marked by point B. 
This only needs to be true for the \emph{last replicated box} where the 
arclengths are the widest. This requirement can be written
\begin{equation}
d \, \sin (\beta) \geq {OB} \, \sin (\alpha + \beta) ~,
\label{condition2}
\end{equation} 
where ${OB} = d - {BC}$, and ${BC}$ can be found using the law of sines: 
% ${BC} = b \frac { \sin (90 - \alpha/2 - \beta) } { \sin (90 - \alpha/2) }$.
${BC} = b \, { \sin (90 - \alpha/2 - \beta) } / { \sin (90 - \alpha/2) }$.
If a $\beta$ value satisfying these conditions can be found the cone is unique. 
Otherwise, the user is asked to either modify their cone parameters or consider a `random' 
cone (described below).

Finally, to complete the cone we apply an equatorial coordinate system to the unique volume, 
orientated such that the cone points in the direction on the sky specified by the minimum 
and maximum RA and Dec requested by the user. Galaxies within the requested range and 
depth are then drawn from the volume and added to the unique cone.

We note that with TAO one can build a unique \emph{full-sky} catalogue by 
constructing 4 unique light-cones with opening angles $90\!\times\!90$ degrees$^2$ 
and with a depth of up to half of the side length of the simulation box. 
These 4 cones can then be combined by the user in post-processing, placing 
the observer at the origin to provide an all-sky view of the theoretical `sky'.

\subsubsection{Random cones} \label{random}

More typically, however, the volume of the desired cone is larger than the
volume of the simulation cube. In this case one can build a random cone. Although
such cones result in the replication of structure, any
periodicity can be mitigated somewhat by using randomisation
techniques \citep[e.g.][]{Blaizot2005} which produce a more realistic
light-cone with pseudo-unique structure (i.e. repeated but non-periodic).

To remove the appearance of periodically repeating structures three
randomisation transformations are applied within the TAO light-cone module: random
rotation, mirroring, and translation of each repeated simulation cube.
\newline \newline \noindent \textbf{Rotation:} The rotation matrix of the principal axis $x$, $y$ and $z$ is given by
\begin{equation}
\begin{bmatrix}
x' \\
y' \\
z' \\
\end{bmatrix}
=
\left[ {\begin{array}{ccc}
   \hat{\mathbf{u}}_x & \hat{\mathbf{v}}_x & \hat{\mathbf{w}}_x \\
   \hat{\mathbf{u}}_y & \hat{\mathbf{v}}_y & \hat{\mathbf{w}}_y \\
   \hat{\mathbf{u}}_z & \hat{\mathbf{v}}_z & \hat{\mathbf{w}}_z \\
\end{array}} \right]
\begin{bmatrix}
x \\
y \\
z \\
\end{bmatrix}
\end{equation}

\begin{equation}
\begin{cases}
\hat{\mathbf{u}}_x = \cos\theta \cos\psi \\
\hat{\mathbf{u}}_y = \cos\theta \sin\psi \\
\hat{\mathbf{u}}_z = -\sin\theta \\
\end{cases}
\end{equation}

\begin{equation}
\begin{cases}
\hat{\mathbf{v}}_x =  -\cos\phi \sin\psi + \sin\phi \sin\theta \cos\psi \\
\hat{\mathbf{v}}_y =  \cos\phi \cos\psi + \sin\phi \sin\theta \sin\psi \\
\hat{\mathbf{v}}_z = \sin\phi \cos\theta \\
\end{cases}
\end{equation}

\begin{equation}
\begin{cases}
\hat{\mathbf{w}}_x = \sin\phi \sin\psi + \cos\phi \sin\theta \cos\psi \\
\hat{\mathbf{w}}_y = -\sin\phi \cos\psi + \cos\phi \sin\theta \sin\psi \\
\hat{\mathbf{w}}_z = \cos\phi \cos\theta \\
\end{cases}
\end{equation}
where $\phi$, $\psi$ and $\theta$ are Euler angles. For the sake of performance
TAO randomly takes values of 0, 90, 180 or 270 degrees.
\newline \newline \noindent \textbf{Mirroring:} Mirroring of the simulation 
volume can be simply achieved by changing an axis
direction. It should be noted that inversion of all of the principal axis in
combination with rotation may result in the original positions in the
simulation cube, so these combinations are excluded from the randomisation
routine.
\newline \newline \noindent \textbf{Translation:} To translate the cube we 
cut the simulation box at a random position along one
axis and move the sliced volume before this position to the end of
the simulation box along of the same axis. This operation doesn't affect
continuous structure in the box because of the periodic boundary condition in
the original simulation.
\newline \newline \indent To illustrate the effects of replication and randomisation we construct two
mock light-cones built with TAO using the milli-Millennium Simulation \citep{Springel2005}
having box side-length $62.5 \ h^{-1} \rm Mpc$. This is shown in
Figure~\ref{replication}.
The first assumes straight replication of the simulation box (left panel),
while the second applies the above random rotations, shifting and mirroring
(right panel).

The differences between the left and right panels of Figure~\ref{replication}
are striking. At a bare minimum, given the visual nature of astronomical
research clearly non-periodic mock catalogues are desirable. But more
importantly, randomisation removes the possibility of unintended replicated
features creeping into statistical applications of the mock. We emphasise
however that spatially-dependent results should never be drawn from mocks on 
scale-lengths larger than the simulation box itself.

\begin{figure}
  \centering
  \includegraphics[scale=0.5]{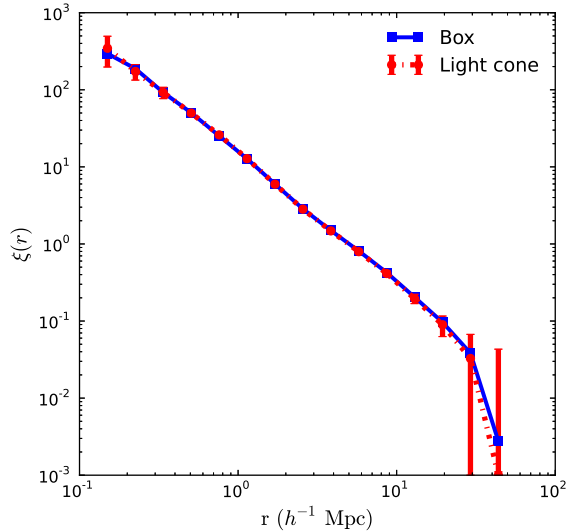}
  \caption{The galaxy 2-point correlation function measured in the original Bolshoi simulation 
  box (blue solid line), and from 20 randomised mock light-cones constructed using the box and 
  the procedures described in Section~\ref{Lightcones} (red dashed line). Error bars represent
  the 3$\sigma$ scatter amongst the 20 cones, and highlight the level of consistancy between 
  the box and cone results.}
  \label{clustering}
\end{figure}

To quantify the effect of such operations on the spatial distribution of galaxies 
along the cone, in Figure \ref{clustering} we plot the real-space 2-point 
correlation function for galaxies drawn from TAO that are more massive than 
$10^{10}\, h^{-1} \rm{M_\odot}$. We first do this for original galaxy positions 
in the the Bolshoi simulation box \citep{Klypin2011}, which has a side-length of 
$250 h^{-1} \rm{Mpc}$ (solid blue line). We then compute the clustering in 20 
cones cut from the same (full) simulation box, each with an area of 
$30 \times 30$ degrees on the sky covering redshift ${0<z<0.2}$ (up to $576 h^{-1} 
\rm{Mpc}$ depth along the line-of-sight, dashed red line). This selection ensures 
that volumes are similar for both cones and the box. Each of the cones 
span six replicated boxes and therefore include all three random transformations described above.
The $3 \sigma$ scatter in the clustering results are used as a 
measure of the clustering uncertainty in the generated cones sample, shown by the 
error bars. Figure \ref{clustering} shows consistent behaviour between the original 
galaxy distribution and randomised cones across the range of scales plotted.
 
\begin{figure}
	\centering
	\includegraphics[scale=0.5]{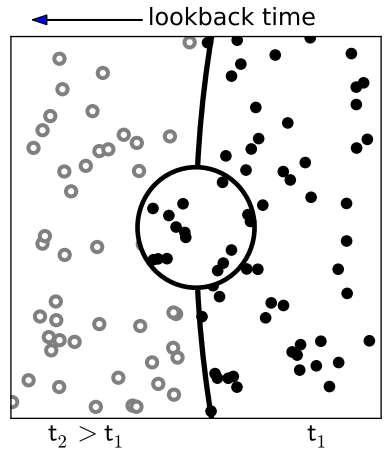}
	\caption{Subhalos from the same parent dark matter halo on a border between
		two redshift zones are kept together. The open dots show the galaxy positions from the
		previous time-step in the simulation, whereas the solid circles show galaxies from the next
		time-step. The large circle in the middle defines the region occupied by one
		particular dark matter halo, which we preserve from splitting when applying time evolution across the cone galaxies.}
	\label{substruct}
\end{figure}

\subsection{Time evolution along the cone}

As we progressively move back through the light-cone away from the point of
observation we see galaxies from earlier and earlier epochs, due to the time
required for the light from each galaxy to cover the distance. Therefore, to
construct an accurate cone we need to not only worry about the spatial
distribution of the galaxies in it, but their evolution as well. Remembering
that a galaxy's radial distance can be directly mapped to a cosmic time, we
place each galaxy in the cone as it appeared in the simulation at the age of 
the Universe corresponding to the distance between the galaxy and the observer.

By doing this we also reduce the consequences of structure replication during
the cone construction processes. Not only will repeated large-scale structures be seen from
different orientations due to the randomisation algorithm described above, but they
are likely to be earlier (or later) versions of their duplicates, perhaps
appearing differently depending on the growth history of each halo--galaxy
system.

There is an additional complication here that is quite subtle yet 
important, as illustrated in Figure~\ref{substruct}. A
key part of the light-cone production is to make sure satellite galaxies are
always connected to their parent dark matter halo. It often happens that a dark
matter halo has its position in the cone very close to the border between
different simulation time-steps. If so, satellite galaxies in that halo may inadvertently 
be split in time across the boundary, and then also be displaced by the
randomisation part of the algorithm. 
In order to provide structural consistency, when building the light-cone, similar to \cite{Carlson2010}, we 
group galaxies (centrals and satellites) by their parent halo association 
as given in the TAO database. Based on the position of the halo centre we then insert these as a unit into the cone,
even if the unit spatially crosses the time boundary at any point. This ensures 
that entire galaxy--halo structures at the same time-step are selected for the cone.

\begin{figure*}
	\begin{center}
		\begin{minipage}{160mm}
			\centering
			\includegraphics[scale=0.51]{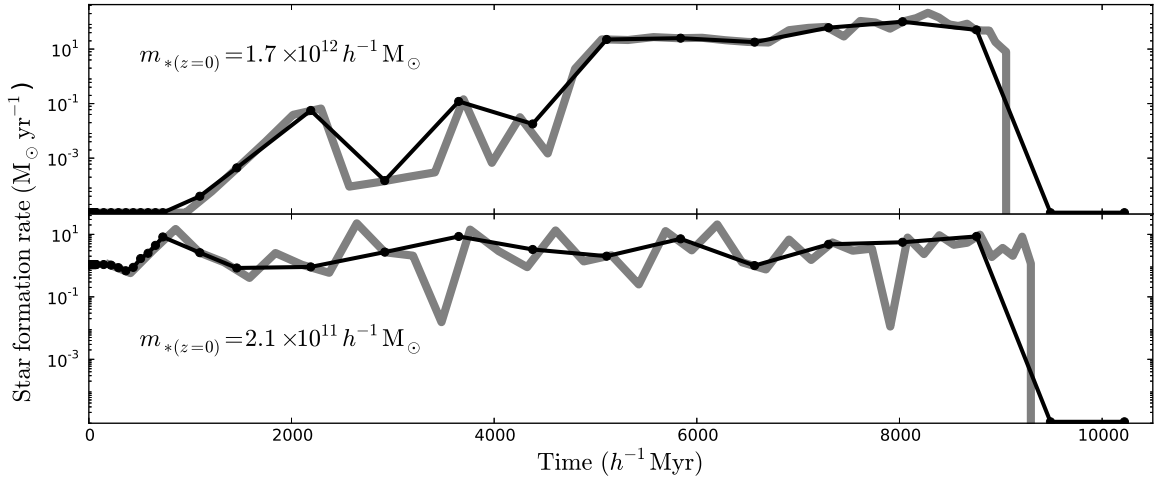}
			\caption{Star formation histories of two galaxies in the TAO database: a large elliptical galaxy 
				(top) and a large Milky Way-type spiral galaxy (bottom).
				For each, the grey thick line shows the original star formation history from the semi analytic
				model on the time grid of the dark matter simulation, whereas the thin black line with points 
				indicates the interpolated star formation history on the grid required by the SED module in TAO. }
			\label{sfh}
		\end{minipage}
	\end{center}
\end{figure*}

A more advanced technique for interpolating the positions of satellite galaxies 
is discussed in \cite{Merson2013}. However, given the discrete nature of simulation 
snapshots we have to accept that all galaxy properties, including positions, 
will reflect only one point in time and are locked to the time-steps of the simulation. 
A more direct way to reduce the effect of discreteness is to run simulations 
with higher time resolution, although this will inevitably result in an increased data size 
that will require more powerful computers with larger storage capacities.

\section{The Spectral Energy Distribution Module}\label{SEDs}

The next crucial step to building mock galaxies is to model their stellar
emission, which further enables the translation of theoretical quantities into mock observables. 
The light emitted by a galaxy is the direct outcome of the formation and
evolution of stars, which is regulated by all the physical mechanisms involved
in galaxy formation.

From a practical point-of-view, modelling galaxy emission can be separated
from the rest of the galaxy formation model. In TAO, this is performed as a 
post-processing step in the spectral energy
distribution module and applied to the TAO galaxy data. This post-processing link
uses the prediction for the star formation rate and metallicity of each 
galaxy at each time step to synthesise galaxy spectral energy 
distributions at the required time in the galaxy history.

Keeping all records of all star formation episodes for every galaxy can
increase the simulation output size significantly. However,
thanks to the relational design of the TAO database we can efficiently 
trace each galaxy history among its progenitors using a system of 
FOF indexes, described above and in \cite{Springel2005}.

Some existing publicly available models offer galaxy luminosities calculated at 
each simulation time step as the model is run, which are then interpolated according to the final position 
in the light-cone \citep{Blaizot2005}. Others offer a more sophisticated interpolation
with k-correction included \citep{Merson2013}. TAO, in contrast, calculates a more accurate
galaxy luminosity directly from spectra constructed `on-the-fly' in post-processing, 
using each star formation history starting from the actual galaxy position in the light-cone. 
This approach also has the advantage of allowing different stellar population synthesis 
models to be tested with existing galaxy data long after the model was originally run.

\subsection{Stellar population synthesis models}

Galaxy light is the superposition of the emission of all the stars in the
galaxy. A galaxy is composed of a series of single stellar populations (SSPs),
i.e. ensembles of stars formed in single episodes with the same age
and metallicity. The SSPs that compose a galaxy either originate in the galaxy
itself through star formation, or are accreted from satellite
galaxies. The emission of every SSP has to be modelled and added to the total
galaxy light. Obviously, as the stars in the galaxy age their emission
changes with time, and the model needs to take this time-dependance into account.

The tools used to accomplish this are stellar population synthesis (SPS)
models, which are libraries of spectra of single stellar populations
built on a grid of ages and metallicities, assuming a particular initial mass
function (IMF) \citep[e.g.][]{BruzualCharlot2003, Maraston2005, ConroyGunnWhite2009}. 
In order to model the galaxy emission as a
function of time, at every time-step in the galaxy model we keep track of all
the single stellar populations in the galaxy, then assign them the corresponding emission
based on their age and metallicity. We then sum over all populations to obtain the
total galaxy light, i.e. its spectral energy distribution (SED). At this
stage the complicating contribution of dust extinction and emission must also be included (see Section~\ref{sec:dust}).

Semi-analytic models are now taking advantage of such methods to generate galaxy light 
\citep{Hatton2003, Tonini2009, Tonini2010, Henriques2011, Merson2013}, which add a level of
sophistication and flexibility. The use of spectra, as opposed to magnitude
tables \citep{DeLucia2004, Croton2006, Baugh2006, Bower2006} brings 
a number of advantages: 1) an increased
precision due to the linear additive nature of luminosity $vs$ the
logarithmic behaviour of magnitudes; 2) an increased accuracy for the
determination of observed magnitudes, which are obtained by integration on
redshifted SEDs rather than through theoretical k-corrections that rely on
toy-model spectra; and 3) an enormous flexibility, introduced when SEDs are modelled
in post-processing. Producing galaxy spectra allows us to separate the
photometric calculations from the semi-analytic model itself, removing the
need to re-run the galaxy model every time we want to change the photometry
specifications. This can include different initial mass functions (IMF), dust
models, filter sets, mock observational errors, and telescope or survey-specific effects.

\begin{figure}
	\centering
	\includegraphics[scale=0.22]{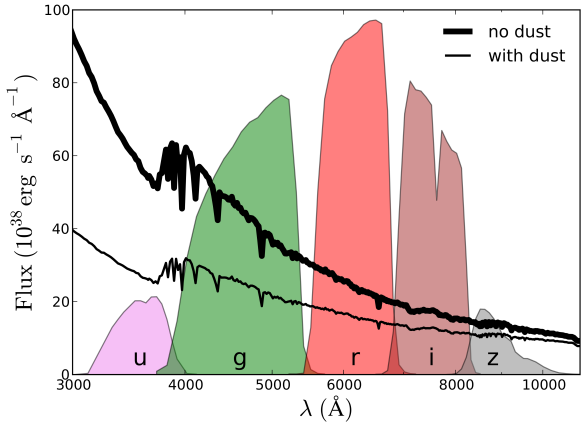}
	\caption{Synthetic spectra of a galaxy from TAO at 
		$z=0$ that has a star formation rate of $15 \rm M_{\odot}\, yr^{-1}$ 
		and stellar mass $2.1 \times 10^{10}\, h^{-1} \rm M_{\odot}$. Here the
		\citet{Maraston2005} SSP model has been assumed. The thick line shows the spectra 
		without dust extinction, while the thin line is the same but with the dust model 
		from \citet{Tonini2012} applied. Over-plotted are the SDSS $u$, $g$, $r$, $i$, $z$ transmission 
		functions using an arbitrary y-axis scale.
	}
	\label{spectra}
\end{figure} 

\subsection{Galaxy star formation histories}\label{star-form-hist}

To calculate a galaxy SED within TAO we require its star formation history, 
defined as the stellar populations present in the galaxy at the time
of observation $\tau_0$, characterised by stellar mass, age, and metallicity. These
populations include stars formed in the galaxy itself and
those that have been accreted from satellites along the merger
tree\footnote{As an aside, the capacity to record the star formation history
of a galaxy in its actual merger tree is an undeniable advantage of
semi-analytic models over many other models techniques: without the
merger tree information toy-models of galaxy evolution can not account for the
complexity of the hierarchical nature of the galaxy assembly, potentially introducing
significant biases \citep{Tonini2012}.}.

We build a two-dimensional age and metallicity grid and collapse onto it the
stars formed across the entire tree up to the point of observation 
(i.e. up to the point $\tau_0$, when we observe the galaxy on the light cone). 
Here, each bin represents a single stellar population of a given age and metallicity, and from the
SPS libraries we select the corresponding spectrum and weight it by the stellar
mass formed. Each of these spectra are then added to the total galaxy light to produce
the final SED.

The challenge in the production of the star formation history is the grid
itself. SPS models are built to follow the vastly different speeds of stellar
evolution. For example, the emission of a young stellar population is dominated by
massive stars and changes on time-scales of $\sim$1 Myr, while an old
stellar population emits steadily on time-scales of $\sim$1 Gyr, with emission
dominated by less massive main-sequence stars. The star formation history grid needs to
provide information on corresponding time-scales in order to produce realistic
galaxy SEDs. In particular, the ultra-violet and optical part of the SED is
heavily (if not entirely) determined by young stellar populations, an issue that
becomes extremely important at high redshifts.

To this end, the star formation histories of every galaxy in the light-cone
must be written by TAO onto a time-varying grid, anchored at the time of observation
$\tau_0$, being finely spaced (steps of 1 Myr) for young ages near $\tau_0$ and
more sparsely spaced towards older ages. Unfortunately, the intrinsic 
semi-analytic model time grid (the snapshots on which the model calculates its 
physics) is not typically spaced like this. For example, the
average time step of a model built using the Millennium Simulation is of the order
of $\sim 300$ Myr. Furthermore, refining a galaxy model grid to the level of precision
required by SPS models is not usually practical due to the huge amount of 
data storage that would be required.

To make up for the loss of information on smaller time-scales we spread the
time-weighted stellar mass produced in each larger semi-analytic model output step onto the
fine SPS time grid close to $\tau_0$. Note that this is equivalent to assuming
that in each simulation time bin we have a constant star formation rate. At the other
end of the SPS time grid the opposite is done, where the mass produced across
multiple model time-steps are collected and re-binned to fill the coarser SPS
grid, accounting for the slow evolution of the old stellar populations.

In Figure~\ref{sfh} we plot two example star formation histories from a
semi-analytic model in TAO, one for a spiral galaxy and one for an elliptical. In both
panels, the \textit{grey line} represents the galaxy star formation rate binned
using the original simulation time grid, summed over all the branches in the
merger tree, while the \textit{black line} represents the star formation history
interpolated over the SPS time grid used in TAO. Notice how, for recent
times (i.e. close to $\tau_0$), the spacing of the SPS grid is very fine, and
the interpolation recovers the SAM star formation history exactly, while for
older ages the spacing becomes sparser, providing an approximated
reconstruction. 

\subsection{Galaxy spectro-photometric properties}

TAO produces galaxy photometry after determining the SEDs. Magnitudes are 
calculated by convolving each galaxy spectrum with a set of
filter transmission functions, which include filters from many of the most commonly
used instruments and surveys. TAO interpolates filters and
spectra on a variable wavelength grid so that the resolution of the
integration is constant no matter the wavelength extension of the filter
function. 

The user is given the choice of both absolute and apparent magnitudes for each filter. In
the case of absolute magnitudes, fluxes are calculated from the total luminosity on
a sphere of radius $R=10 \rm pc$. For apparent magnitudes, the flux is
calculated on a sphere of radius equal to the luminosity distance corresponding
to the redshift of the galaxy. The spectrum is dimmed and stretched in
wavelength according to the redshift, and then convolved with each selected filter
function. Notice that this operation is exact, as opposed to the
approximation of using a k-correction, as both the rest-frame and observed
(stretched) spectra are known.

In Figure~\ref{spectra} we show an example SED from a $z=0$ star forming 
galaxy in TAO (thin line), where the \cite{Maraston2005} SPS was chosen.
It is evident that recent and more intense star formation increases the
stellar emission in the UV and optical wavelengths, and the young stellar
populations dominate the bolometric luminosity. 
Also shown is the effect of including the contribution of dust
extinction (in the UV/optical) and emission (mid-to-far infrared; thin line). 
The inclusion of dust in the theoretical spectra is available to the user 
as an option within the TAO SED science module, which we now discuss. 

\subsection{Dust}
\label{sec:dust}

Dust emission and extinction plays a fundamental role in shaping the galaxy SEDs,
especially in cases of significant star formation, which is particularly
relevant at high redshifts. 
In TAO, we provide dust modelling as a separate step and allow the user to 
choose whether to apply dust to the galaxy SED or not. Two popular models are 
available in TAO and are described below.

\subsubsection{Slab model}

In the slab model, stars and dust are assumed to be 
homogeneously distributed in an infinite plane-parallel slab with the same 
vertical scale. The dust-attenuated luminosity at 
wavelength $\lambda$, ${\rm L}{\mathstrut}^{\rm obs}_{\lambda}$, is given by
\begin{equation}
{\rm L}{\mathstrut}^{\rm obs}_{\lambda} = {\rm L}{\mathstrut}^{\rm intr}_{\lambda} \: \cfrac{1 - e^{-\tau_{\lambda}^{\rm eff} \sec{i}}}{\tau_{\lambda}^{\rm eff} \sec{i}},
\end{equation}
where ${\rm L}{\mathstrut}^{\rm intr}_{\lambda}$ is the intrinsic luminosity of 
the disc and $i$ is its inclination angle. We define the effective dust 
opacity as $\tau^{\rm eff}_{\lambda} = (1-\omega_{\lambda})^{\scriptscriptstyle 1/2} (1+z)^{\scriptscriptstyle -1/2} \: \tau_{\lambda}$, 
where $\tau_{\lambda}$ is the face-on dust opacity. Following \citet{Devriendt1999}, $\tau_{\lambda}$ 
is expressed as a function of the neutral hydrogen column density of the 
disc, N$_{\rm H}$:
\begin{equation}
\tau_{\lambda} = \left(\cfrac{A_{\lambda}}{A_V}\right)_{\rm Z_{\odot}} \left(\cfrac{Z}{Z_{\odot}}\right)^{s} \left(\frac{N_{\rm H}}{2.1 \times 10^{21} \: {\rm atoms\ cm}^{-2}}\right).
\label{eq:tau}
\end{equation}
Here $(A_{\lambda} / A_V)_{\rm Z_{\odot}}$ is the extinction curve for solar 
metallicity ${\rm Z_{\odot}}$ (see \citealt{Mathis1983}) and varies with gas metallicity 
${\rm Z}$ and wavelength such that $s = 1.35$ for $\lambda > 2000$\AA, and $s = 1.6$ for 
$\lambda >2000$\AA{} (see \citealt{Guiderdoni1987} for more details). 

The first term of the dust opacity, $(1-\omega_{\lambda})^{1/2}$, accounts for 
scattering effects, where $\omega_{\lambda}$ is the albedo. The second term of 
Eq. \ref{eq:tau}, often used in semi-analytic models (e.g. \citealt{Kitzbichler2007}, 
\citealt{Garel2012}), introduces an additional scaling of the dust-to-gas ratio with 
redshift which is in broad agreement with observational trends seen in high 
redshift galaxies, e.g. \citet{Reddy2006}.

\subsubsection{Calzetti prescription}

The dust content of a galaxy can be parameterised with the colour excess
$E(B-V)$, which is defined as a re-normalisation of the spectrum. Physically, 
dust content is associated with the presence of Type II supernovae,
which are the main contributors to the metals that constitute the dust grains.
Such grains are short-lived (with a life-span of the order of $\sim 10-100$
Myr), so it is sensible to associate the dust content with the instantaneous
star formation rate in the galaxy (see \citealt{Tonini2011}):
\begin{equation}
E(B-V) = R_{dust}A\cdot \left(\frac{\dot{m}_{*}}{\dot{m}_{*,0}}\right)^{\gamma}+B~.
\end{equation}
Here, $A = (e^{3}-e^{-2})^{-1}$, $B = -Ae^{-2}$,
$\dot{m}_{*,0} = 1.479\,M_{\odot} \rm{y}^{-1}$, and $\gamma=0.4343$ are values obtained
with a calibration of the GOODS sample discussed in \citet{Daddi2007} and the
Andromeda galaxy. We use a Calzetti extinction curve, which produces absorption 
blue-wards of the Johnston K band, and re-emission red-wards (see 
\citealt{Calzetti1997} and \citealt{Calzetti2001}).

\section{The Mock Image Module} \label{mockimage}

In addition to the spectral energy distribution of individual galaxies or 
other objects, many astronomical instruments take images of the wider sky 
in selected parts of the electromagnetic spectrum using broad or narrow filters. 
Having the ability to model such images with synthetic data provides an 
important link in understanding how galaxy properties are connected with their 
observation.

However, fairly comparing images made from different simulations and models can
be problematic, and the size of such datasets poses additional challenges for image
generation. Consistency when making images is essential to minimise artificial
differences due to the underlying simulation data and processing. With this in
mind, TAO is able to produce mock telescope images in a consistent,
seamless, and user friendly way.

The TAO image generation module uses the SkyMaker software package 
(see \citealt{SkyMaker} for further details).
This software takes data produced by the light cone and SED modules and creates 
realistic telescope images that include many of the usual observational effects, 
like telescope aperture, optical defects, sky characteristics, and aureole around
bright galaxies.

TAO users can request multiple images per submitted job.  Each image requires 
the following parameters: 
\begin{itemize}
	\item the desired bandpass filter out of those chosen in the SED module, 
	\item the observed magnitude limits of galaxies to include, 
	\item the RA and Dec coordinates to centre the image on, 
	\item the opening angles of the image within the light-cone area, 
	\item the redshift range of galaxies to include, and
	\item the resolution of the final output image file. 
\end{itemize}

\begin{figure*}
	\begin{center}
		\centering
		\includegraphics[scale=0.55]{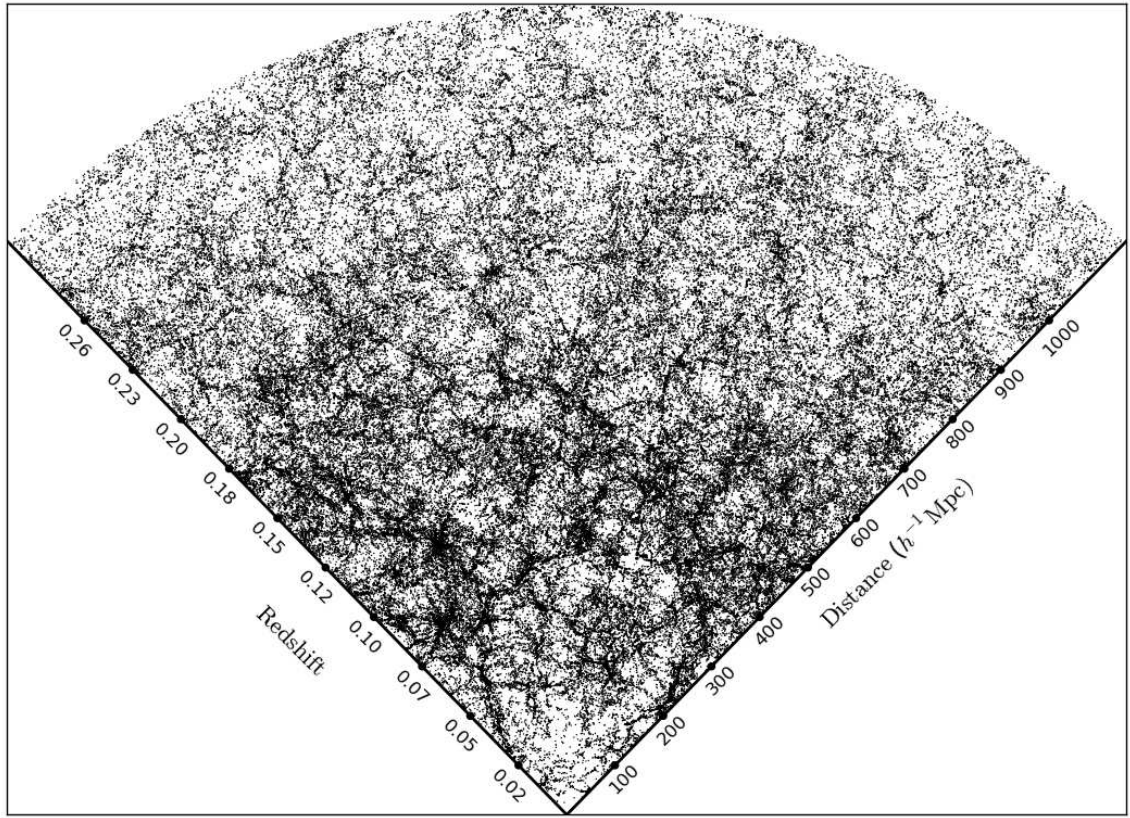}
		\caption{A local SDSS-type galaxy mock catalogue with a limiting magnitude of $r < 17.77$, constructed 
			using TAO as described in Section~\ref{SDSS}.}
		\label{sdss-wedge}
	\end{center}
\end{figure*}

To construct an image the module requires information
about galaxy positions in the light-cone, galaxy morphology, and total 
magnitudes in the requested filters.
From the bulge and the disk properties SkyMaker renders galaxies,
using a de Vaucouleurs profile for the bulge and an exponential profile for the
disk. Each galaxy is then then placed in the field-of-view to build up the final image.

The point spread function model used in SkyMaker is a convolution of the
following components: atmospheric blurring for ground-based instruments\footnote{Atmospheric 
blurring is not currently used in TAO since the default instrument is the Hubble 
Space Telescope, however ground-based instruments will be added in a subsequent update.}, 
telescope motion blurring, instrument diffraction and aberrations, optical diffusion effects, 
and intra-pixel response. Sky background, noise, saturation, and quantisation 
modelling are also added by SkyMaker to reflect the inherent noise and artefacts 
in charge-coupled devices (CCDs).

Some parameters, like image pixel size and average filter wavelength, are calculated by
TAO during job processing. Others are pre-chosen to produce the best image quality; 
in Table~\ref{mock-images-params} we list these. Any further parameters not listed 
are kept at their default value, as assumed in the SkyMaker software package 
\citep{SkyMaker}. As the TAO system develops we will add further customisation 
options to the user interface for more precise image generation control.

\begin{table}
\centering
\begin{tabular}{|l|c|}
\hline
\textbf{Parameter} & \textbf{Value} \\
\hline
\hline
Exposure time & 300 sec \\
Magnitude zero-point & 26 mag \\
Background surface brightness & 50 $\rm mag\,arsec^{-2}$ \\
Range covered by aureole & 50 pixels \\
Aureole surface brightness & 16 $\rm mag\, arsec^{-2}$ \\
PSF oversampling factor & 7 \\
PSF mask size & 512 pixels \\
Diameter of the primary mirror & 3.5 m \\
Diameter of the secondary mirror & 1 m \\
\hline
\end{tabular}
\caption{SkyMaker default settings in the Mock Image Module initial TAO release. 
With time these will be added to the module interface for control by the user.}
\label{mock-images-params}
\end{table}

\section{Examples of Mock Catalogues} \label{applications}

\begin{figure*}
	\centering
	\includegraphics[scale=0.39]{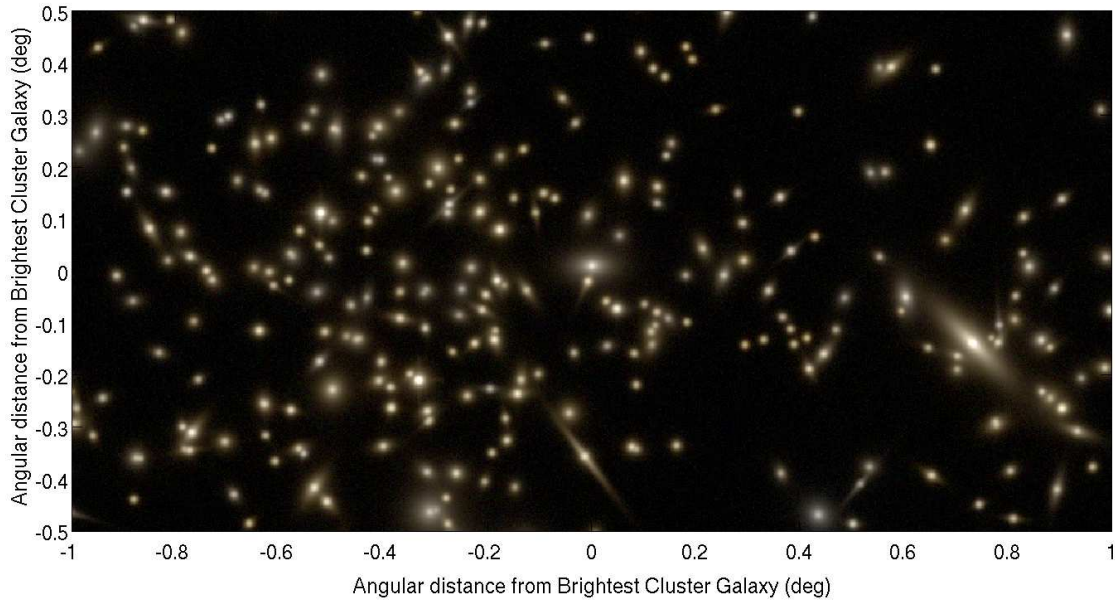}
	\caption{An SDSS-$gri$ composite image of a $M_{vir} = 5.71 \times 10^{14}\, h^{-1} \rm M_{\odot}$ 
		galaxy cluster at $z = 0.08$, within a 2x1 degree field-of-view, rendered using 
		the TAO image module as described in Section~\ref{image}. This particular cluster 
		contains 119 galaxies brighter than a limiting galaxy magnitude of $r=17.6$.}
	\label{cluster-image}
\end{figure*}

There are many ways in which TAO mock catalogues can be built using the tools 
described so far. One example is the work of \cite{Duffy2012} to make predictions 
for the Australian Square Kilometre Array Pathfinder (ASKAP) telescope neutral 
hydrogen surveys of WALLABY and DINGO. TAO comes with an expanding set of popular 
survey presets, which can be run as-is or used as a template for more specific 
requirements. To illustrate the utility of TAO, in this section we outline what we expect is a 
common use case, to build a representation of the local universe as a light-cone and then image
a particularly massive cluster of galaxies within its borders.
 
\subsection{SDSS mock catalogue light-cone} \label{SDSS}

One of the largest observational extragalactic databases to date is the Sloan
Digital Sky Survey (SDSS) catalogue\footnote{http://www.sdss3.org/dr9/scope.php} 
\citep{Abazajian2003,sdss-dr9}. The SDSS main catalogue covers approximately
14000 square degrees of the sky across a redshift range {$0 < z \ \simlt \
0.4$}.
Many thousands of papers have been written using SDSS data, making it the
highest impact repository for extragalactic science in the history of
astronomy. Hence, the SDSS is a natural place to start if one wants
to construct mock analogues of the local universe.

To reconstruct an SDSS volume using TAO we perform the following steps:
\begin{itemize}
	\item General Properties: We select a light cone geometry, then a simulation and
	galaxy formation model. Here, the Millennium simulation \citep{Springel2005}
	and SAGE galaxy model (Croton et al. submitted) are adopted. 
	The RA and Dec opening angles are chosen
	to be 90 and 60 degrees, respectively, with a redshift range of $0 < z < 0.3$.
	\item Spectral Energy Distribution: For this particular cone we select a BC03
	stellar population model assuming a Chabrier initial mass function. All
	filters marked with `SDSS' are included for output -- $u$, $g$, $r$, $i$, $z$
	-- both apparent and absolute. We apply the Calzetti dust model described in 
    \cite{Tonini2012}.
	\item Selection: Our cone is chosen to be volume limited by including all
	galaxies with stellar masses greater than $10^{8}\, h^{-1} \rm M_\odot$.
\end{itemize}

After processing, TAO sends an email to the user with a link to the constructed 
light cone for download. This particular cone contains 3,122,823 galaxies. In
Figure~\ref{sdss-wedge} we plot these galaxies to illustrate their spatial
distribution. The catalogue also contains as many galaxy properties as predicted 
by the model and requested by the user. Furthermore, the exact same cone can be 
reconstructed using a different model in the database, or with a different 
underlying simulation. This provides the user with an estimate of the theoretical 
uncertainty between different models and simulations.

\subsection{Imaging a local galaxy cluster} \label{image}

We can use the SDSS light-cone constructed above to extend our exploration of the 
local galaxy volume beyond the spectroscopic. This can be done by employing the 
TAO image module to predict what an imaging survey might see within the cone. 

To this end, from the history page we download the SDSS light-cone data to a local 
machine and identify the most prominent galaxy cluster within its volume. We define 
this to be the dark matter halo with the largest galaxy membership (central+satellites) 
brighter than a limiting magnitude of 17.6 in the $r$-band (but other selections 
are of course possible). The winning halo was found at $z=0.08$, had a virial mass 
of $5.71 \times 10^{14}\, h^{-1} \rm M_{\odot}$, and contained 119 galaxies. We note 
the angular position of the cluster centre and its extent on the sky, both in RA and Dec.

Once identified, the mock image module can be used to build a realistic SDSS-$gri$ 
visualisation to be compared with observations. Within the TAO interface, we 
revisit the SDSS light-cone data through the history page. From the options 
presented we choose to image the cone. We then centre our image on the centre of 
the cluster, select a field-of-view of 2x1 degrees to capture its surrounding environment, 
and take the full redshift depth of the cone, {$0.0 {<} z {<} 0.3$}, which allows us 
to model the effect of interlopers. This is done individually for each of the $g$, $r$, 
and $i$ filters to produce three images. We keep most of the SkyMaker imaging 
options fixed at their default values, with the exception of matching the SDSS 
exposure time of 53.9 seconds, and choosing not to populate the image with foreground 
stars. 

After processing by TAO we are again notified via email that our results 
are ready. The individual $g$, $r$ and $i$ images are then downloaded and 
combined to form a single composite image using the ds9 package. This image 
is shown in Figure ~\ref{cluster-image}.

\section{Summary} \label{summary}

In this paper we have described a new cloud-based virtual laboratory, the Theoretical
Astrophysical Observatory (TAO), that enables any astronomer to produce mock galaxy 
catalogues based on selectable combinations of a dark matter simulation, semi analytic 
galaxy formation model, and stellar population synthesis model. TAO is built in a modular
fashion, starting with the simulation and model database, that flows upward
through a series of science modules to connect to the user via a simple web
interface.

The key features of TAO are:
\begin{itemize}
\item A flexible design with an intuitive web interface that allows public access to 
many dark matter simulations, galaxy models, and stellar population synthesis
models;
\item Advanced techniques to create custom mock light-cones of large volumes with
realistic structure;
\item Galaxy-by-galaxy spectral energy distribution modelling obtained in post-processing, 
providing accurate galaxy photometry using galaxy star formation and metallicity 
histories;
\item The ability to image light-cone data using the SkyMaker image processing package, 
integrated into the TAO workflow.
\end{itemize}

The flexibility built into TAO make it useful for many applications. Some examples include:
\begin{itemize}
\item Making survey predictions and planning observing strategies;
\item The comparison of observational data with simulations and models;
% \item Testing how different physical prescriptions in the same galaxy
% model affect galaxy evolution;
\item The comparison of different galaxy models run on the \textit{same} dark matter
simulation;
\item The comparison of a single galaxy model run on \textit{different} dark matter
simulations;
\item Exploring the effects of different stellar population synthesis models
and dust extinction prescriptions on a galaxy's photometric evolution;
\item Generating mock images and testing source finding algorithms.
\end{itemize}

TAO is an open source project and can be freely deployed and further developed
by members of the community. The Swinburne TAO portal can also be used by simulators 
and modellers to make selected data of importance available to the public, which is especially useful when the
resources to do this in-house are not available or expensive to implement.
Enabling the community with free access to state-of-the-art theoretical data 
facilitates data reuse and multiplies its value.

\section*{Acknowledgements}

The authors would like to thank Alistair Grant, Jarrod Hurley, Gin Tan, Jennifer Piscionere, and Manodeep Sinha. We 
also appreciate the time given by the many astronomers who tested TAO during 
its initial development. Special thanks goes to Andrew Benson for his help importing 
the Galacticus semi-analytic model into the TAO database. 
DC acknowledges receipt of a QEII Fellowship by the Australian Research Council 
(DP1095506). SM, AD and GP are funded from the ARC Laureate Fellowship grant of 
S. Wyithe (FL110100072).

TAO is part of the All-Sky Virtual Observatory and is funded and supported by 
Astronomy Australia Limited, Swinburne University of Technology, and the 
Australian Government. The latter is provided though the Commonwealth's Education 
Investment Fund and National Collaborative Research Infrastructure Strategy, 
particularly the National eResearch Collaboration Tools and Resources (NeCTAR) 
project. TAO was constructed as a collaboration between between Swinburne 
University of Technology and Intersect Australia and is hosted on the gSTAR 
national facility at Swinburne.

The Millennium Simulation was carried out by the Virgo Supercomputing Consortium 
at the Computing Centre of the Max Plank Society in Garching. It is also publicly 
available at http://www.mpa-garching.mpg.de/Millennium/.
The Bolshoi Simulation was carried out by A.~Klypin, J.~Primack and S.~Gottloeber 
at the NASA Ames Research Centre. The simulation and data products can additionally 
be found at http://astronomy.nmsu.edu/aklypin/Bolshoi/.
The Semi-Analytic Galaxy Evolution (SAGE) model used in this work is a publicly 
available codebase that runs on the dark matter halo trees of a cosmological 
N-body simulation. It is available for download at https://github.com/darrencroton/sage.

Finally, the authors would like to thank the anonymous referee, whose careful 
reading of this paper resulted in many valuable improvements.

\bibliographystyle{mn2e} 
\bibliography{./reference}

\end{document}